# GIP-RAG: An Evidence-Grounded Retrieval-Augmented Framework for Interpretable Gene Interaction and Pathway Impact Analysis

Fujian Jia[1], Jiwen Gu[1], Cheng Lu[2], Dezhi Zhao[1], Mengjiang Huang[3], Yuanzhi Lu[4], Xin Liu[1,§], Kang Liu[1,§]

1, Kanghua Juntai Biotech Co. Ltd., Room 1504, Building 7, Dongwang Jingyuan, No. 768 Jingwang Road, Kunshan, Suzhou, Jiangsu Province, China

2, Department of Radiation Oncology, Cancer Treatment Center, The Second Affiliated Hospital of Hainan Medical University, Haikou, China

3, Department of Nutrition, University of California, Davis, USA

4, Department of Pathology, The First Affiliated Hospital of Jinan University, Tianhe Qu, Guangzhou, China.

§, Correspondence:

Xin Liu, xin_liu@kanghuajuntai.com

Kang Liu, kellen0101@live.com

**Abstract**

Understanding mechanistic relationships among genes and their impacts on biological pathways is fundamental for elucidating disease mechanisms and advancing precision medicine. Although numerous public databases provide extensive information on molecular interactions and signaling pathways, effectively integrating heterogeneous knowledge sources and performing interpretable multi-step reasoning across biological networks remains challenging.

Here, we present GIP-RAG (Gene Interaction Prediction through Retrieval-Augmented Generation), a computational framework that combines biomedical knowledge graphs with the reasoning capabilities of large language models (LLMs) to infer and interpret gene interactions. The framework first constructs a unified gene interaction knowledge graph by integrating curated interaction data from multiple public resources, including KEGG, WikiPathways, SIGNOR, Pathway Commons, and PubChem. Given user-specified genes, a query-driven subgraph retrieval module dynamically extracts relevant evidence from the knowledge graph. The retrieved subgraphs are incorporated into a structured prompting strategy that guides LLM-based stepwise reasoning to identify direct and indirect regulatory relationships and generate mechanistic explanations supported by biological evidence. Beyond pairwise interaction inference, this network further introduces a pathway-level functional impact assessment module that simulates the propagation of gene perturbations through signaling networks and evaluates potential pathway state alterations. Evaluation across multiple biological scenarios demonstrates that the framework produces consistent, interpretable, and evidence-supported insights into gene regulatory relationships.

Overall, GIP-RAG provides a general and interpretable paradigm for integrating biological knowledge graphs with retrieval-augmented large language models to facilitate mechanistic reasoning in complex molecular networks.

**Introduction**

In recent years, the systematic analysis of gene interaction networks has attracted substantial attention in biomedical research [1–3]. Such networks constitute a fundamental basis for understanding the regulatory mechanisms of complex biological systems, elucidating disease pathogenesis, and guiding precision medicine. Although public databases such as KEGG [4], WikiPathways [5], SIGNOR [6], Pathway Commons [7], and PubChem [8] provide abundant information on molecular interactions and biological pathways, efficiently integrating multi-source, heterogeneous data and performing mechanistic inference between genes while preserving interpretability remains a major challenge in computational biology.

Existing pathway- or network-based methods for gene interaction inference typically rely on graph-theoretic analyses, path-search algorithms, or statistical association models, which can to some extent reveal direct or local regulatory relationships [9,10]. However, their flexibility and interpretability are limited when addressing multi-hop, cross-pathway, and context-dependent indirect regulatory relationships [11–12]. Meanwhile, with the rapid development of large language models (LLMs) and their remarkable reasoning capabilities demonstrated in natural language processing tasks, their potential applications in biomedical knowledge integration and complex relational inference have increasingly drawn attention[13-15]. Nevertheless, inference based solely on the generative capacity of LLMs often suffers from incomplete knowledge coverage and hallucination effects, lacking stringent biological evidence constraints, which limits their reliability for high-confidence mechanistic inference[16].

To address these challenges, we propose a retrieval-augmented generation (RAG)-based gene interaction reasoning framework, termed GIP-RAG (Gene Interaction Prediction through RAG). This framework systematically integrates high-confidence biological knowledge from multiple public databases to construct a unified gene interaction knowledge graph. A query-driven subgraph retrieval strategy is then employed to explicitly introduce structured evidence into the LLM reasoning process. On this basis, carefully designed structured prompt engineering enables evidence-grounded, step-by-step reasoning and mechanistic interpretation. Furthermore, GIP-RAG is extended to pathway-level functional impact assessment, simulating the propagation of gene perturbations through signaling networks and characterizing potential pathway state alterations from a systems biology perspective.

In summary, the main contributions of this study are as follows:
1. Construction of a high-quality, multi-source gene interaction knowledge graph, together with subgraph retrieval and structured reasoning interfaces, providing a reliable evidence foundation for mechanistic gene–gene interaction inference;
2. Proposal of the GIP-RAG framework, which combines RAG techniques with LLMs to achieve evidence-based and interpretable gene interaction reasoning;

3. Development of a pathway-level functional impact assessment module that simulates signaling network reconfiguration under gene perturbation hypotheses, offering a practical tool for systems biology research.

Overall, this framework provides a general and interpretable technical paradigm for integrating biological knowledge graphs with the reasoning capabilities of large language models, and offers potential value for precision medicine, disease mechanism analysis, and novel therapeutic target discovery.

## Methods

We propose a computational framework termed GIP-RAG (Gene Interaction Prediction through Retrieval-Augmented Generation), which integrates multi-source biomedical knowledge graphs with the advanced reasoning capabilities of large language models (LLMs) to infer and explain potential mechanistic interactions among user-specified genes.

The core strength of GIP-RAG lies in its tight coupling of precise retrieval from structured biological knowledge with semantic and logical reasoning by generative models, enabling interpretable inference of direct or indirect gene–gene relationships supported by existing biological evidence.

### 1. Overall Framework Design

The GIP-RAG framework consists of four major stages:

- Multi-source data integration: Extraction and normalization of biological pathway and chemical interaction data from five public databases.

- Biomedical knowledge graph construction: Transformation of standardized interaction data into a unified knowledge graph repository.
- Query-driven subgraph retrieval: Dynamic retrieval of evidence subgraphs relevant to the queried genes.
- LLM-based structured reasoning and synthesis: Conversion of retrieved evidence into biologically coherent mechanistic explanations through prompt engineering.

**2. Data Sources and Knowledge Integration**

To comprehensively capture gene interaction relationships across multiple biological layers, we systematically collected pathway and interaction data from the following widely used, community-curated or expert-reviewed public databases:

- KEGG: Provides manually curated molecular interaction and biochemical reaction networks;
- WikiPathways: A community-maintained resource covering diverse biological pathways;
- SIGNOR: A signaling database emphasizing causal and directional regulatory relationships;
- Pathway Commons: An integrated collection of pathway and molecular interaction data aggregated from multiple authoritative sources;
- PubChem: Contains gene–compound interactions and associated bioactivity annotations.

Each database was independently parsed to extract both pathway-level and molecular-level interaction information. Particular emphasis was placed on interactions with explicit directionality and mechanistic annotations, including but not limited to activation, inhibition, binding, phosphorylation, and complex formation.

## 3. Data Standardization and Knowledge Graph Construction

To enable seamless integration of heterogeneous data sources, all extracted interactions underwent a unified standardization and harmonization pipeline, as described below.

### 3.1. Format Standardization

All raw interaction records were converted into a standardized triplet format (source entity, interaction type, target entity).Each record additionally retained rich metadata, including source entity, target entity, interaction type, pathway context, originating database, and evidence annotations. Only high-confidence interactions supported by manual curation or experimental evidence were retained.

### 3.2. Gene Identifier Harmonization

All gene names were mapped to HGNC-approved official gene symbols, ensuring consistent gene representation across databases.

### 3.3. Semantic Normalization

Heterogeneous interaction descriptors were mapped to a controlled vocabulary to ensure semantic consistency across sources (e.g., "activates" and "positively regulates" were unified under the term "activation").

### 3.4. Confidence Assessment

Each interaction was assigned a composite confidence score based on:

- The number of independent databases supporting the interaction;
- The evidence level of the original source (manual curation, literature support, or inference);
- The completeness of directionality and mechanistic annotation.

Only interactions exceeding a predefined confidence threshold were included in downstream knowledge graph construction.

### 3.5. Gene Interaction Knowledge Graph Repository

The standardized interactions were stored in a graph database (GraphStore), forming a biomedical knowledge graph composed of multiple node and edge types. The schema is defined as follows:

- Node types:
  - Gene nodes (represented by HGNC symbols),
  - Pathway nodes,
  - Compound nodes.

- Edge types:
  - Gene–gene regulatory edges,
  - Gene–pathway association edges,
  - Gene–compound biochemical interaction edges.

Each edge is enriched with attributes including interaction type, source database, confidence score, and literature references. The graph structure is designed to support efficient graph traversal and subgraph extraction.

### 4. Query Processing and Subgraph Retrieval

Given an input list of query genes, the system performs the following steps to retrieve relevant evidence:

1. Gene identifier validation: Input genes are standardized and validated using the same harmonization pipeline employed during graph construction.

2. Dynamic subgraph traversal: The subgraph retrieval module performs depth-first or breadth-first searches to dynamically extract gene-centered evidence subgraphs. The traversal depth ( D ) is adaptively controlled to balance biological relevance and computational complexity. Retrieved subgraphs include direct interactions, shared regulators, and indirect associations mediated through pathway nodes.

## 5. Structured Reasoning and Explanation Generation Based on Knowledge Subgraphs

Retrieved subgraphs are provided as external knowledge inputs to the LLM, and a multi-stage prompt engineering strategy is employed to guide evidence-based logical reasoning.

### 5.1. Structured Design of Reasoning Prompts

Prompt templates follow a hierarchical structure to ensure clarity of instruction and consistency of input:

- Role and task definition: The LLM is explicitly instructed to act as a molecular biology expert tasked with inferring functional gene relationships based on provided interaction evidence.

- Structured contextual input: Subgraph information is supplied in JSON format, including query genes, edge lists (with interaction types and data sources), and aggregated confidence metadata.

- Chain-of-thought reasoning instructions: A step-by-step reasoning strategy is enforced, comprising:

    1. Path identification: Identification of all direct and multi-hop paths connecting the query genes;

    2. Evidence evaluation: Integration of interaction types, database authority, multi-source consistency, and functional module coherence;

3. Synthesis and ranking: Aggregation of high-confidence paths to infer regulatory directionality or functional association, followed by ranking according to evidence strength.

- Output specifications: The LLM is required to generate structured yet natural-language explanations that explicitly include core conclusions, supporting mechanistic descriptions (with cited evidence sources), qualitative assessments of evidence strength, and stated limitations.

**5.2. Pathway Function Impact Assessment: Mechanistic Evaluation Based on Pathway Perturbation**

Building upon inferred gene relationships, we further developed a pathway-level functional impact assessment module to evaluate the potential effects of gene perturbations (e.g., overexpression, knockdown, or mutation) on overall pathway states.

1. Pathway-context-enhanced retrieval: Expanded subgraph extraction is performed with particular emphasis on key input/output nodes, regulatory hubs, feedback loops, and pathway cross-talk points.

2. Prompt framework for functional impact reasoning: A system-biology-oriented prompt is designed as follows:

   – Input: Enhanced pathway subgraphs combined with hypothetical gene perturbation scenarios (e.g., "loss of function of gene A").

   – Reasoning steps:

   a. Local impact simulation: Simulation of signal propagation originating from the perturbed node;

   b. Network-level impact assessment: Identification of downstream functional

modules most strongly affected;

  c. Mechanistic analysis: Explicit consideration of compensatory mechanisms, redundant pathways, and feedback failures to enable fine-grained impact evaluation.

– Output: A structured report describing altered core pathway states, mechanistic cascades, system-level adaptive responses, and a high-level summary of the resulting global signal network reconfiguration pattern.

## 6. RAG-LLM System Implementation and Deployment

To enable scalable retrieval-augmented reasoning over the constructed biomedical knowledge graph, we implemented a modular RAG-LLM infrastructure that integrates graph retrieval, semantic indexing, and large language model inference within a unified computational pipeline.

### 6.1 Component #1 Knowledge Graph Storage and Query Layer

The standardized interaction data were stored in a graph database environment to enable efficient traversal and subgraph extraction. The knowledge graph was implemented using Neo4j, which supports property graph models and high-performance graph queries. Nodes represent biological entities (genes, pathways, compounds), and edges represent directional biological interactions annotated with interaction type, evidence source, and confidence scores. Graph queries were executed through Cypher query language, allowing dynamic extraction of gene-centered subgraphs with adjustable traversal depth during runtime. The retrieved graph structures were serialized into structured JSON objects to serve as contextual inputs for downstream RAG reasoning.

## 6.2 Component #2 Semantic Retrieval and Indexing Module

To support efficient contextual retrieval, the graph-derived textual and structured knowledge units were indexed using an embedding-based semantic retrieval system. Biological interaction records and pathway descriptions were first transformed into textual evidence blocks. These blocks were embedded using a transformer-based sentence embedding model (Bio-domain optimized embedding models such as BioBERT or general embedding models depending on deployment constraints). The embeddings were stored in a vector database (FAISS), enabling fast approximate nearest-neighbor search for semantically relevant biological evidence.

During query execution, user-specified genes were used to generate retrieval queries that combined: graph-based subgraph extraction and embedding-based semantic similarity search. The retrieved evidence blocks were then aggregated and passed to the reasoning stage.

## 6.3 Component #3 RAG Orchestration Pipeline

The retrieval-augmented generation pipeline was implemented using a modular orchestration framework (LangChain). This pipeline coordinates the following steps:

**1.** Query preprocessing**:** Gene identifiers provided by users are normalized to HGNC-approved gene symbols.

**2.** Graph-based evidence retrieval**:** The graph traversal module retrieves candidate interaction paths and pathway contexts.

**3.** Evidence ranking and filtering**:** Retrieved interactions are ranked according to confidence scores, database authority, and redundancy across multiple sources.

**4.** Context assembly**:** High-confidence interaction paths and supporting metadata are converted into structured JSON prompts.

**5.** LLM reasoning and synthesis**:** The assembled evidence context is fed into the LLM together with structured reasoning instructions.

This design ensures that the LLM operates under explicit evidence constraints rather than relying on implicit knowledge contained within model parameters.

6.4 Component #4 Large Language Model Reasoning Engine

The reasoning engine was implemented using a large language model deployed through a scalable inference environment. Depending on computational infrastructure, the model can be hosted through cloud-based APIs or locally deployed model servers. The LLM was configured with controlled generation parameters to balance reasoning depth and output stability, including temperature control, token limits, and deterministic decoding settings when necessary.

The entire pipeline was implemented in Python, integrating widely adopted libraries for graph processing, vector retrieval, and LLM orchestration. The modular architecture enables flexible replacement of individual components (e.g., embedding models or LLMs) without altering the overall framework.

This implementation allows GIP-RAG to efficiently integrate structured biological knowledge with the reasoning capabilities of modern language models, providing an extensible infrastructure for interpretable gene interaction inference.

**Results**

**1. Overview of the RAG-Driven Gene Interaction Inference Framework**

We first constructed and validated a gene interaction inference framework based on Retrieval-Augmented Generation (RAG), designed to infer potential regulatory relationships between user-specified genes under the support of multiple public biological databases. The framework integrates manually curated pathway and molecular interaction information from KEGG, WikiPathways, SIGNOR, Pathway Commons, and PubChem. Through a unified knowledge representation and retrieval strategy, it enables efficient utilization of heterogeneous biological knowledge.

Across the overall test set, the system consistently returned structured inference results that include interaction types, upstream–downstream directionality, biological pathway context, and supporting literature evidence. Compared with generative models that rely on a single database or lack retrieval augmentation, the proposed framework demonstrated clear advantages in inference consistency, interpretability, and result completeness, providing a robust foundation for the analysis of complex gene regulatory networks.

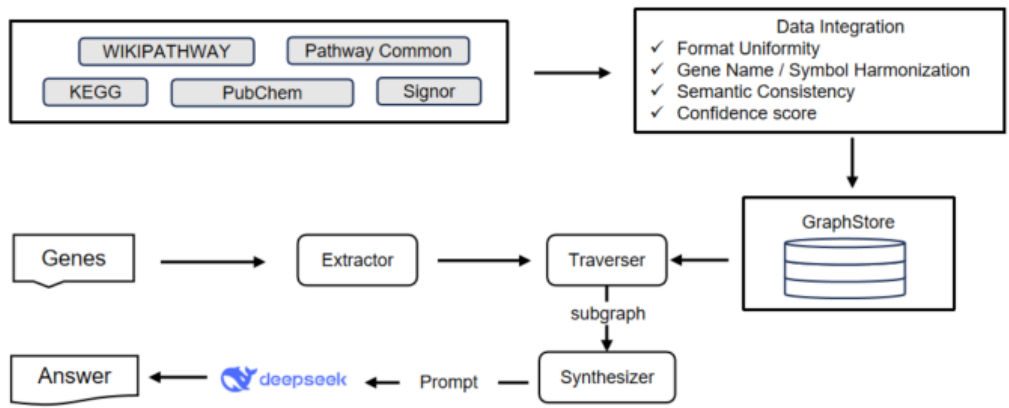

**Figure 1.** Schematic overview of the RAG-driven gene interaction inference framework.

**2. Multi-Level Validation of Gene Interaction Type and Directionality Inference**

Based on the query processing and subgraph retrieval as well as structured reasoning procedures described in the Methods section, we systematically evaluated the framework's ability to infer gene interactions across multiple levels of biological complexity. The evaluation design strictly corresponded to four representative biological scenarios outlined in the Methods, enabling a comprehensive assessment of model performance under varying evidence conditions.

**2.1 Inference of Direct Regulatory Relationships Within the Same Pathway**

For gene pairs located within the same pathway and linked by well-defined direct regulatory relationships (e.g., phosphorylation or transcriptional activation), GIP-RAG accurately retrieved edges representing these direct interactions along with their complete attributes. The reasoning module then generated explanations that explicitly described the biochemical mechanism, regulatory direction, and original database references. In all tested cases, such interactions were correctly inferred, and the inferred regulatory directions were fully consistent with source database annotations. These results confirm the framework's core capability to retrieve high-confidence direct evidence and produce precise mechanistic interpretations.

> **Example 1.1**
>
> **Input Genes:** EGF, EGFR
> **Output:**
> Conclusion: EGF → (directly activates) → EGFR Regulatory Relationship Confirmed (clear upstream-downstream relationship)
> Evidence:
> 1. Localization Verification (Key Upstream Node):
>    * **KEGG Pathway (EGFR tyrosine kinase inhibitor resistance):** Explicitly documents that `EGF` has an activating effect on `EGFR` (`activation:-->`) and lists `EGF` as one of the initiating signaling molecules of this pathway.
>    * **WikiPathway (EGFR tyrosine kinase inhibitor resistance):** Explicitly records `EGF` as the "ligand activation" upstream molecule of `EGFR`.
>    * In all relevant pathway data (Signaling by EGFR, EGFR tyrosine kinase inhibitor resistance), `EGF` consistently appears as the initiating ligand signal, while `EGFR` is its directly interacting first-level transmembrane receptor.
>
> 2. Interaction Analysis (Direct Activation and Binding):
>    * **KEGG:** Relationship type is `PPrel` (protein-protein relation), with the subtype `activation:-->`.
>    * **WikiPathway:** The action type is "ligand activation."
>    * **PathwayCommons:** The action type is `controls-state-change-of`, indicating that EGF controls the state change of EGFR.
>    * **SIGNOR:** Provides direct experimental evidence (`DIRECT: YES`), with the mechanism being `binding` and the effect being `up-regulates activity` of EGFR, accompanied by a specific literature PMID (`12297050`).
>
> Comprehensive Analysis:
> As a ligand, `EGF` specifically binds to its receptor `EGFR` on the cell membrane, directly inducing its dimerization and autophosphorylation of the tyrosine kinase domain, thereby transitioning it from an inactive to an active state. This "ligand-receptor" activation event is the core initiating step of the EGFR signaling pathway. It is consistently and thoroughly described across multiple authoritative pathway databases and is supported by direct experimental evidence. All data are consistent and unequivocally support the regulatory relationship in which EGF is upstream of EGFR and directly activates its activity.

```
Example 1.2
Input Genes: MYC, MAPK8
Output:
MAPK8 → (Direct Phosphorylation and Activation) → MYC Regulatory Relationship Confirmed (Lv1)
Basis and Analysis:
1. Localization Validation
MAPK8 (JNK1): In multiple databases, including Signor, the KEGG MAPK signaling pathway, and the KEGG Wnt signaling pathway, MAPK8 is identified as a core kinase of the MAPK signaling pathway. Typically activated by upstream MAP2K4/7, it functions as a protein kinase responsible for phosphorylating a series of downstream transcription factors (e.g., JUN, ELK1), representing a classic key node in signal transduction.
MYC: In the KEGG Wnt signaling pathway, KEGG PI3K-Akt signaling pathway, and Signor databases, MYC is consistently defined as a terminal transcription factor in signaling pathways. It is regulated by upstream signals (e.g., CTNNB1, LEF1) at the transcriptional level and/or by post-translational modifications (e.g., phosphorylation), thereby driving phenotypes such as cell proliferation and growth.
2. Interaction Analysis
Evidence for Direct Action: The Signor database provides high-quality direct experimental evidence.
Record: ENTITYA: MAPK8, ENTITYB: MYC, EFFECT: up-regulates activity, MECHANISM: phosphorylation, PMID: 10551811, DIRECT: YES.
Interpretation: This record clearly indicates that MAPK8 directly upregulates MYC activity via a phosphorylation mechanism. PMID 10551811 refers to specific experimental research literature, and it is annotated as a direct interaction (DIRECT: YES), indicating the highest confidence level.
Pathway Logic Support: The Wikipathway MAPK signaling entry contains a regulatory chain: "MAPK8 → JUN → Proliferation". As MYC is a well-known pro-proliferation transcription factor that functions synergistically with JUN, the evidence from Signor directly links MAPK8 to MYC, refining this logical pathway.
Consistency and Corroboration: While other databases (KEGG, wikipathway) do not directly list the MAPK8→MYC relationship, they all position MAPK8 downstream in kinase cascades and MYC at the transcriptional output terminal. This is entirely consistent with the directionality of the direct regulatory relationship discovered in Signor, and no contradictory evidence exists.
……
```

**Figure 2.** Representative results of direct regulatory relationship inference within the same pathway.

## 2.2 Inference of Indirect Regulatory Relationships Within the Same Pathway

For gene pairs residing within the same pathway but functionally connected through one or more intermediate nodes, the framework successfully extracted multi-hop paths connecting the genes via dynamic subgraph traversal with an adjustable depth parameter (D). The path identification and evidence evaluation steps embedded in the reasoning prompts guided the LLM to integrate these indirect paths into a coherent mechanistic narrative (e.g., "Gene A indirectly activates Gene C by inhibiting an intermediate suppressor, Gene B"). The results demonstrate that the framework can effectively identify and explain indirect regulatory relationships mediated by shared regulators or signaling cascades, thereby overcoming the limitations of approaches focused solely on direct interactions.

> **Example 2.1**
>
> **Input Genes:** EGFR, BRAF
> **Output:**
> ……
> 3. Upstream-Downstream Relationship Determination
> According to the hierarchical conclusion criteria:
> - **Clear Upstream-Downstream Relationship** (meeting any two conditions):
>   - ☑ **Located at a key upstream node:** EGFR is the initiating receptor of the pathway, and BRAF is a downstream kinase.
>   - ☑ **Interaction type is direct activation/inhibition:** Although there is no direct interaction between EGFR and BRAF, HRAS→BRAF in the indirect path is a direct activation (experimentally verified).
>   - ☑ **Existence of experimentally verified evidence:** SIGNOR, KEGG, etc., provide experimentally supported indirect pathways (e.g., PMID: 7706312 for HRAS-BRAF).
> - **Conclusion:** EGFR acts as the upstream gene, activating BRAF via an indirect pathway.
> **Hierarchical Result:**
> **EGFR → (Indirectly Activates) → BRAF Regulatory Relationship Confirmed (Lv1)**
> **Evidence:**
> - KEGG "MAPK signaling pathway" (PATHWAY: MAPK signaling pathway) shows that EGFR activates BRAF via the GRB2-SOS1-HRAS cascade.
> - WikiPathway "MAPK signaling pathway" (Pathway: MAPK signaling pathway) clearly annotates EGFR as upstream and BRAF as downstream.
> - SIGNOR "EGFR signaling" (PATHWAY: EGFR signaling) provides experimental evidence for HRAS directly activating BRAF (direct interaction), and EGFR activating HRAS via GRB2-SOS1.
> - No contradictory evidence: All databases consistently support the upstream position of EGFR. BRAF mutations may independently activate the MAPK pathway but do not alter the upstream-downstream relationship in the normal pathway.
>
> ⚠ **Wet-lab Validation Suggestion:** Since the regulation is indirect, it is recommended to validate the effect of EGFR on BRAF phosphorylation through gene knockout/inhibitor experiments (e.g., detecting changes in BRAF phosphorylation levels following EGFR inhibition).
> ……

> **Example 2.2**
>
> **Input Genes:** HRAS, MAP2K1
> **Output:**
> **Analysis Conclusion**
> HRAS → (Indirect Activation) → MAP2K1 Regulatory Relationship Confirmed (Lv1)
> **Basis:**
> 1. **Consistent Pathway Hierarchy**: In KEGG pathways such as MAPK signaling pathway, EGFR tyrosine kinase inhibitor resistance, and PI3K-Akt signaling pathway, the standard signaling flow is: HRAS → RAF (ARAF/BRAF/RAF1) → MAP2K1. This indicates that HRAS is a clear upstream signaling molecule of MAP2K1.
> 2. **Direct Interactions and Experimental Validation:**
> Direct Activation of Downstream by HRAS: The SIGNOR database (high-quality signaling relationships), within the context of EGFR signaling and PI3K-Akt signaling pathway, clearly records that HRAS directly binds to and upregulates the activity of BRAF (EFFECT: up-regulates activity, MECHANISM: binding, DIRECT: YES, PMIDs: 7706312, 18098337, 21779497).
> Direct Upstream Activation of MAP2K1: The KEGG database explicitly records in the aforementioned pathways that BRAF/RAF1/ARAF directly activates MAP2K1 through phosphorylation (Relation_type: PPrel, subtypes: activation:-->;phosphorylation:+p).
> 3. **Clear Regulatory Cascade**: Integrating evidence from multiple databases forms a complete, classic RAS-RAF-MEK (MAP2K1)-ERK signal transduction cascade: HRAS (in its GTP-activated form) recruits and activates RAF family kinases (BRAF/RAF1), and the activated RAF subsequently phosphorylates and activates MAP2K1 (MEK1).
> ……

**Figure 3.** Examples of indirect gene interaction inference based on pathway context.

## 2.3 Cross-Pathway Gene Interaction Inference

To evaluate the framework's ability to integrate cross-pathway knowledge, we tested gene pairs belonging to different biological pathways but potentially involved in functional crosstalk. The query-driven subgraph retrieval module successfully constructed cross-pathway subgraphs by leveraging pathway nodes or shared compound nodes. Based on these subgraphs, the reasoning module inferred plausible regulatory relationships (e.g., a gene product in one pathway acting as a signaling molecule influencing another pathway) and explicitly identified key hub nodes that bridge the two pathways. This capability highlights the potential of GIP-RAG to discover functional associations that extend beyond individual pathway boundaries through integrated knowledge graph reasoning.

---

**Example 3.1**

**Input Genes:** PIK3CA, TP53
**Output:**
......

**Analysis Conclusion**
PIK3CA → (Indirectly Inhibits) → TP53 Regulatory Relationship Confirmed (Lv1)
......

**2. Interaction Analysis and Evidence Chain Construction**
The confirmation of the regulatory relationship is based on the following high-confidence chain of direct interactions:
**Core Evidence Chain: PIK3CA -> AKT -> MDM2 -> TP53**

1. **Activation from PIK3CA to AKT (Core Pathway Logic):**
   - Mechanism: `PIK3CA`, as the PI3K catalytic subunit, activates upon stimulation, leading to PIP3 generation. PIP3 recruits and activates PDK1 and AKT.
   - Evidence Sources: This step is a defining event in the PI3K/AKT pathway, extensively documented in `pubchem_relations`, `kegg_relations` (`PI3K-Akt signaling pathway`), `wikipathway_relations` (`PI3K-Akt signaling pathway`), and `signor_relations` (`PI3K-Akt signaling pathway`) with entries describing `PI3K` generating `PIP3` and activating `AKT`. For example:
     - `signor_relations`: "PI3K (complex) ... up-regulates quantity ... PIP3 (smallmolecule)" (PMID: 12040186, 24647478).
     - `signor_relations`: "PIP3 (smallmolecule) ... up-regulates activity ... AKT (proteinfamily)" (PMID: 23119004, 23633519).

2. **Direct Activation (Phosphorylation) from AKT to MDM2:**
   - Mechanism: Activated AKT phosphorylates the MDM2 protein (e.g., at residues S166, S188), enhancing its E3 ubiquitin ligase activity and nuclear localization, thereby promoting p53 degradation.
   - Evidence Sources:
     - `pubchem_relations` (PI3K/AKT Signaling): Equation: "ATP + Oncoprotein Mdm2 ? ADP + p-S166,188-MDM2";  Control: "activated by Active AKT". This reaction equation and control condition clearly describe the phosphorylation and activation of MDM2 by AKT.
     - `kegg_relations` (PI3K-Akt signaling pathway): "AKT3 ... MDM2 ... activation ... phosphorylation".
     - `signor_relations` (PI3K-Akt signaling pathway): "AKT (proteinfamily) ... up-regulates activity ... MDM2 (protein) ... phosphorylation ... PMID: 11259415".
   **(Supported by experimental verification PMID)**

3. **Direct Inhibition (Ubiquitination-Mediated Degradation) from MDM2 to TP53:**
   - Mechanism: MDM2, a key negative regulator of p53, ubiquitinates p53 and targets it for proteasomal degradation, thereby inhibiting p53 protein stability and transcriptional activity.
   - Evidence Sources:
     - `kegg_relations` (PI3K-Akt signaling pathway): "MDM2 ... TP53 ... inhibition".
     - `signor_relations` (PI3K-Akt signaling pathway): "MDM2 (protein) ... down-regulates quantity ... TP53 (protein) ... ubiquitination ... PMID: 10523848".
   **(Supported by experimental verification PMID)**
......

> **Example 3.2**
>
> **Input Genes:** MET, RAF1
> **Output:**
> MET → (Indirect Activation) → RAF1 Probable Upstream-Downstream Relationship (Lv2)
> **Basis:**
> **1. Co-Localization and Hierarchy Analysis in Pathways:**
> **Pathway Confirmation:** MET and RAF1 are co-localized in the EGFR signaling network and related pathways. Both are listed as key nodes in the "EGFR signaling" pathway within the SIGNOR database.
> **Node Hierarchy:** In the canonical receptor tyrosine kinase (RTK)-RAS-MAPK signaling axis, MET (the receptor) is positioned at the most upstream end, receiving ligand (e.g., HGF) signals; RAF1, as a key kinase in the MAPK pathway, is located downstream of RAS proteins. Therefore, based on pathway topology, MET is an upstream regulatory factor of RAF1.
> **2. Analysis of Interaction Relationships and Regulatory Pathways:**
> **Lack of Direct Interaction:** Across all provided database records, no direct physical interaction or direct regulatory relationship (e.g., phosphorylation) was found between MET and RAF1.
> **Clear Indirect Regulatory Path:** Evidence supports that MET indirectly activates RAF1 via the classical GRB2-SOS-RAS signaling module.
> The SIGNOR database explicitly records: "HRAS → (up-regulates activity via binding) → BRAF" (PMID: 21779497). RAF1 and BRAF belong to the same RAF family of kinases and are functionally equivalent in this signaling context, sharing the same activation mechanism.
> Although the databases do not directly show a MET → HRAS entry, the KEGG "EGFR tyrosine kinase inhibitor resistance" pathway clearly illustrates MET activating the downstream GRB2-SOS1-HRAS cascade. This mechanism is entirely consistent with how other RTKs like EGFR activate RAS. The wikipathway entry also supports MET activating NRAS (functionally similar to HRAS) via SOS.
> **Experimental Evidence Support:** The relationship of HRAS activating BRAF/RAF1 in SIGNOR is backed by direct experimental validation (PMID: 21779497), providing high-confidence support for this key link in the indirect pathway.
> **Conclusion Derivation:** Integrating pathway hierarchy and validated molecular interactions, the most probable regulatory path is: MET (activation) → GRB2/SOS (guanine nucleotide exchange factor activation) → HRAS (GTP-bound activation) → RAF1 (binding and activation). Therefore, MET is an upstream indirect activator of RAF1.
> ……

**Figure 4.** Inference results for gene interactions across distinct biological pathways.

## 2.4 System-Level Network Inference with Multi-Gene Inputs

Finally, we evaluated the framework's performance with multiple (>2) input genes. The system first performed dynamic subgraph traversal to retrieve an interaction network encompassing all input genes. The reasoning module then conducted not only pairwise inference of potential relationships between genes but, more importantly, system-level integrative analysis, identifying core regulatory hubs, potential functional modules, and dominant signal flow directions within the network. This capability provides a powerful computational tool for investigating the coordinated behavior of gene sets (e.g., differentially expressed gene sets) under specific physiological or pathological conditions.

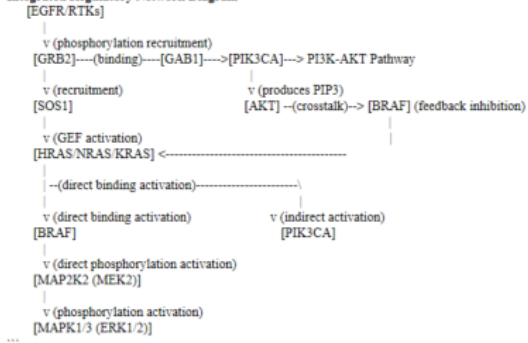

**Figure 5.** System-level regulatory network inference with multi-gene inputs.

## 3. Pathway-Level Functional Impact Assessment Results

Using the pathway functional impact assessment module developed in this study, we simulated and evaluated the cascading effects of functional perturbations in specific genes (e.g., knockdown, overexpression, or gain-of-function mutations) on associated pathway networks.

Given a hypothetical gene perturbation (e.g., "loss of activity of Gene A"), the module first performed pathway context–enhanced retrieval to obtain an extended subgraph containing key input/output nodes and feedback loops. Subsequently, guided by the functional impact reasoning prompt framework, the LLM simulated signal propagation from the perturbed node along the edges of the subgraph. The results show that the framework can clearly identify direct first-order downstream targets and correctly infer the direction of activity changes (upregulation or downregulation) based on interaction types (activation or inhibition).

Beyond local effects, the framework further assessed the global consequences of perturbations across the network. Specifically, it successfully identified significantly affected downstream functional modules (e.g., cell cycle or apoptosis), potential compensatory or redundant pathways that may be activated to maintain network stability when primary pathways are disrupted, and alterations in feedback loops that could lead to system oscillations or steady-state imbalance.

The inference outputs were presented in the form of structured reports, detailing affected core pathway states, mechanistic cascade events, and ultimately summarizing the global network reconfiguration patterns (e.g., an overall shift from pro-survival signaling to pro-apoptotic signaling).

### Example 5.1

**Input:** EGFR loss of function

**Output:**

**Core Affected Pathways and Functional Status**

**1. EGFR Tyrosine Kinase Inhibitor Resistance Pathway**
- Overall Functional Status: Functional Reprogramming / Bypass-Driven Signal Dysregulation
- Key Gene Alterations and Hierarchy:
  - Inactivated Gene: EGFR (Upstream receptor node).
  - Compensatory Enhancement of Bypass Receptor Gene Function: MET, IGF1R, AXL, FGFR3, IL6R, etc. (Parallel upstream nodes). Data indicate these receptors can activate downstream adaptor proteins and kinases (JAK1, SRC, GAB1, PLCG1, SHC2) shared with EGFR via PPrel (activation:-->) relationships.
- Inferred Logic:
  - Direct Inhibition: EGFR inactivation directly severs its own activating signals to JAK1, SRC, GAB1, PLCG1, SHC2 (e.g., EGFR, JAK1, PPrel, activation:-->).
  - Bypass Activation: The network contains numerous parallel inputs. For example, receptors like MET, IGF1R, and AXL can activate the exact same set of downstream nodes (MET, JAK1, PPrel, activation:-->; IGF1R, SRC, PPrel, activation:-->, etc.). These bypass receptors are unaffected by EGFR status and can initiate downstream signaling upon ligand (HGF, IGF1, GAS6, etc.) binding, achieving "circumvention" of the EGFR signaling axis.
  - Signal Output Reshaping: The source of activation for downstream core signaling hubs (STAT3, AKT3, MAPK1) shifts from "EGFR-dominant" to "multi-receptor parallel input". This may alter signal persistence and confer intrinsic or acquired resistance to EGFR-targeted monotherapy.

**2. PI3K–Akt Signaling Pathway**
- Overall Functional Status: Functionally Active (Maintained via Synergy of Bypass and Negative Feedback Release)
- Key Gene Alterations and Hierarchy:
  - Weakened Direct Input: EGFR inactivation reduces activation of P3R3URF-PIK3R3 (PI3K complex) via GAB1 and SHC2 (KEGG: EGFR,GAB1,PPrel,activation; SHC2,GRB2...; GAB1,P3R3URF-PIK3R3, PPrel,activation).
  - Activation of Alternative Inputs: Nodes such as RAC1 (TLR2,RAC1,PPrel,activation), CD19 (SYK,CD19,PPrel,activation), and PTK2 (integrin signaling) can directly activate PI3K (RAC1,P3R3URF-PIK3R3,PPrel,activation, etc.).
  - Potential Release of Negative Regulation: EGFR inactivation may weaken indirect regulation of PTEN (no direct interaction in data). More critically, inhibitory phosphorylation of RAF1 by AKT3 (AKT3,RAF1,PPrel,inhibition:--;phosphorylation:+p) may diminish due to fluctuations in AKT activity, releasing its cross-inhibition of the MAPK pathway.
- Inferred Logic:
  - Despite the weakened EGFR input, network redundancy is extremely high. Multiple signaling axes parallel to EGFR (e.g., integrin-FAK, BCR-SYK, GPCR-RAC) can directly activate PI3K, ensuring PIP3 generation and AKT3 membrane localization/activation are not completely shut off.
  - Phosphorylation of AKT3 by PDPK1 and mTORC2 (Sieg PubChem, SIGNOR) is a key activation step; these nodes are not under direct EGFR control. As long as upstream lipid signals exist, baseline AKT activity can be maintained.
  - AKT's downstream pro-survival (inhibition of BAD, CASP9, FOXO3), pro-proliferative (inhibition of CDKN1A/B, GSK3B), and metabolic regulatory functions persist, supporting cell survival.

......

### Example 5.2

**Input:** BRAF gain of function

**Output:**

**Core Affected Pathways and Functional Status**

Based on the gain-of-function activation (constitutive kinase activity) of the BRAF gene (typically the V600E mutation) and its central position in the signaling network, the functional status of the specified pathways is inferred as follows:

**1. MAPK Signaling Pathway**
- Overall Functional Status: Functionally Active, with Signal Dysregulation (Broken Negative Feedback, Sustained Signaling).
- Key Gene Alteration and Hierarchical Position:
  - BRAF: A core signaling node (MAPKKK). Its constitutive activation directly and persistently phosphorylates and activates downstream MEK (MAPKK).
- Inferred Logic:
  - BRAF's constitutive activity leads to sustained activation of the MEK-ERK cascade, bypassing regulation by upstream growth factor receptors (e.g., EGFR).
  - Failure of Negative Feedback Loops: Normally, ERK activation leads to feedback inhibition of upstream signaling components via phosphorylation, such as upregulating DUSPs (dual-specificity phosphatases) to dephosphorylate ERK itself or phosphorylating SOS and RAF family proteins (e.g., CRAF) to inactivate them. However, mutant BRAF (especially V600E) is relatively insensitive to some upstream negative feedback (e.g., inhibitory phosphorylation of CRAF), leading to reduced efficacy of this feedback loop and sustained signal output.
  - Signal Reprogramming: The signal flow shifts from the normally regulated "Growth Factor-RAS-RAF-MEK-ERK" mode to a constitutively active "Mutant BRAF-MEK-ERK" mode, partially decoupling upstream and downstream regulation.

......

**Pathway Compensation / Bypass / Feedback Changes**

**1. Bypass Activation (Bypass Signaling)**

Mechanism: The BRAF mutation itself constitutes a "downstream bypass" of upstream EGFR signals. This is the core mechanism of EGFR TKI resistance. Similarly, MEK inhibitors may fail due to feedback reactivation of upstream RTKs (e.g., EGFR, IGF1R) or compensation by parallel pathways (e.g., PI3K-Akt).

**2. Parallel Pathway Compensation**

Mechanism: The PI3K-Akt pathway acts as a key compensatory pathway. When the MAPK pathway is strongly inhibited by BRAF inhibitors (e.g., vemurafenib), tumor cells frequently reactivate the PI3K-Akt pathway via feedback reactivation or upregulation of upstream RTKs (e.g., EGFR, PDGFR), thereby maintaining critical survival signals and leading to resistance. In the BRAF mutation context, MAPK and PI3K-Akt pathways are already in a co-activated state. Inhibition of either pathway alone may allow the other to provide sufficient survival and proliferative signals.

......

> **Example 5.3**
>
> **Input:** Functional Activation of HRAS and MAP2K1
> **Output:**
> **I. Core Affected Pathways and Functional Status**
> **MAPK Signaling Pathway**
> Key Gene Alterations: HRAS (Upstream GTPase node), MAP2K1/MEK1 (Core signal transduction node).
>
> **Overall Functional Status:** Functional Activation (Inferred Strength: High). The core signal flow of this pathway (MEK → ERK) is persistently and forcibly initiated. The classical dependence on growth factor receptors (e.g., EGFR) is weakened, leading to signaling that is constitutive and autonomous. Sustained phosphorylation of downstream ERK will drive its nuclear translocation, strongly regulating transcription and cellular functions.
> **PI3K–Akt Signaling Pathway**
> Key Gene Alterations: HRAS (Cross-activation node).
>
> Overall Functional Status: Functional Reprogramming / Signal Perturbation (Inferred Strength: High). The activation of this pathway decouples from complete dependence on classical RTKs (e.g., EGFR, IGF1R), shifting to HRAS-dominated input. Simultaneously, the strongly activated MAPK pathway can indirectly limit further RTK input to PI3K through various negative feedback mechanisms (e.g., phosphorylation-mediated inhibition of RTKs, IRS1, etc.). Consequently, the maintenance of PI3K-Akt signaling relies more heavily on the direct action of HRAS, establishing a bypass-driven activation state that is partially uncoupled from classical RTK input.
>
> **III. Pathway-Level Compensatory / Bypass / Feedback Alterations**
> **Bypass Activation (Key Mechanism):**
> The activation of HRAS/MAP2K1 itself constitutes a perfect bypass to EGFR-targeted therapy. When EGFR is inhibited by drugs, the constitutive activation of these downstream nodes can completely circumvent the blocked upstream signaling, reactivating MAPK and partially PI3K signals, leading to primary or acquired resistance to EGFR TKIs.
> **Parallel Pathway Compensation:**
> The MAPK and PI3K-Akt pathways form a potent compensatory alliance in their downstream functions. Even if one pathway is partially inhibited (e.g., using a MEK inhibitor), the other pathway (e.g., HRAS-sustained PI3K-Akt signaling) can still independently drive crucial pro-survival, anti-apoptotic, and metabolic signals, resulting in limited efficacy of monotherapy. This is the primary rationale for combination targeted therapy (e.g., MEKi + PI3Ki/Akti).
>
> **IV. Summary of Global Signaling Network Remodeling Patterns**
> Integrating the above analyses, the dual functional activation of HRAS and MAP2K1 triggers the following system-level remodeling of the signaling network:
> **Remodeling Pattern:** Exhibits characteristics of "downstream node dominance + multi-pathway synergistic imbalance." The signaling driver core shifts downward and becomes fixed at the HRAS-MAP2K1 hub module, away from membrane receptors (e.g., EGFR).
> **Pathway Synergy:** Not all pathways are equally activated. Instead, a synergistic imbalance pattern emerges: "MAPK pathway forced activation" serves as the engine, "PI3K-Akt pathway bypass synergy" as the booster, "EGFR pathway feedback inhibition" as the consequence, and "Wnt pathway functional reshaping" as the adaptation. Oncogenic pathways (MAPK, PI3K) are synergistically activated, while certain tumor-suppressive functions (e.g., potential tumor-suppressive roles of canonical Wnt in some contexts, normal growth regulation represented by EGFR dependence) are suppressed or reprogrammed.
> **Signal Flow:**
> Upstream Input: Trends towards diversification and desensitization. Due to strong negative feedback, the network's response threshold to various growth factors (EGF, FGF, etc.) increases, and dependence decreases.
> Core Processing: Trends towards centralization. Signals converge on two potent nodes: HRAS-MAP2K1-ERK and HRAS-PI3K-Akt.
> Downstream Output: Trends towards redundancy. For key functional modules like proliferation, survival, and metabolism, the outputs of MAPK and PI3K-Akt pathways are highly overlapping, creating functional redundancy that enhances network robustness (i.e., resistance to therapy).
> **Therapeutic Implications:** This remodeled network possesses intrinsic resistance to EGFR TKIs and exhibits a high probability of compensatory resistance to single-pathway targeted therapy (e.g., MEK inhibitors). Effective intervention strategies must simultaneously target the two synergistic cores, MAPK and PI3K-Akt, while considering their profound impact on cellular metabolism and survival signaling. This network state is commonly observed in subsets of melanoma, lung cancer, and squamous cell carcinomas, representing a paradigmatic example of "oncogenic signal fixation" and "feedback escape" in systems biology.

**Figure 6.** Representative example of pathway-level functional impact assessment induced by gene perturbation

**Discussion**

In this study, we propose GIP-RAG, a gene interaction inference framework based on retrieval-augmented generation (RAG), which aims to enable interpretable inference of potential mechanistic relationships between genes through the systematic integration of multi-source biomedical knowledge graphs and the reasoning capabilities of large language models (LLMs). Compared with traditional gene regulatory network inference methods that rely on statistical correlations or black-box models, this framework emphasizes logic-based reasoning constrained by existing biological evidence, thereby partially addressing the long-standing challenges of limited interpretability and verifiability in inferred gene interactions.

The introduction of the RAG architecture represents a core design element of this work. Its key advantage lies in explicitly incorporating external structured knowledge into the reasoning process of language models, which effectively mitigates the "hallucination" issues commonly observed in purely generative models. Within this framework, the language model does not perform inference in isolation but instead conducts stepwise reasoning under the constraints imposed by retrieved high-confidence subgraph evidence. Moreover, by explicitly constraining the reasoning steps through structured prompt engineering, the model is guided to perform mechanistic analysis in a predefined logical sequence rather than merely producing declarative conclusions.

We further extend gene interaction inference to pathway-level functional impact assessment by introducing a pathway impact evaluation module that analyzes the potential effects of gene functional perturbations on global signaling networks. This module provides a systems-level mechanistic perspective by simulating signal propagation, identifying key regulatory nodes, and characterizing potential compensatory pathways. As a result, the framework is capable not only of addressing whether "gene A influences gene B," but also of inferring through which pathways such effects may occur and how they may directionally alter overall pathway states. This design facilitates a network-level understanding of regulatory reprogramming in complex biological processes, particularly in scenarios characterized by multi-gene coordination and extensive pathway crosstalk.

In addition to its methodological contributions, the GIP-RAG framework may provide practical value in clinic practice especially in precision oncology and multidisciplinary tumor board (MDT) decision-making. Modern cancer treatment increasingly relies on molecular profiling, where multiple genetic alterations are simultaneously detected through next-generation sequencing. However, interpreting the functional relationships among these alterations and understanding their combined impact on signaling pathways remains a major challenge for clinicians. Individual gene mutations rarely act in isolation; instead, they function within complex regulatory networks involving pathway crosstalk, feedback regulation, and compensatory signaling mechanisms.

In this context, systematic pathway-level interpretation of gene interactions is particularly important for MDT discussions. By integrating knowledge from multiple curated pathway databases and performing evidence-grounded reasoning over gene interaction networks, the proposed framework enables structured interpretation of complex multi-gene relationships. Such analyses may help identify key regulatory hubs, clarify whether multiple mutations converge on the same signaling pathway, and distinguish driver alterations from secondary or context-dependent events. Furthermore, pathway-aware reasoning may support the rational design of combination targeted therapies. Many targeted drugs act on different nodes within the same signaling cascade or across interconnected pathways. Understanding how gene perturbations propagate through signaling networks can provide insights into potential synergistic drug combinations or mechanisms of resistance. Therefore, frameworks such as GIP-RAG that integrate biological knowledge graphs with interpretable reasoning may serve as a useful computational aid for MDT teams when evaluating complex genomic profiles and exploring mechanistically informed therapeutic strategies.

Despite these strengths, several limitations of the present study should be acknowledged. First, the inference capability of GIP-RAG is inherently dependent on the coverage and annotation quality of existing public databases; consequently, its performance remains limited for gene relationships that are insufficiently studied or lack robust supporting evidence. Second, heterogeneity across databases with respect to evidence grading, interaction definitions, and biological context may still influence inference outcomes, despite mitigation through confidence assessment and multi-source integration. In addition, the current framework primarily focuses on qualitative and mechanistic reasoning and does not incorporate quantitative expression data or dynamic parameters, which limits its ability to model regulatory strength or precise changes in pathway activity. Furthermore, the reasoning behavior of language models remains sensitive to prompt design and model scale, and different configurations may lead to variations in inferred details. This highlights the necessity of more systematic robustness evaluations and comparative analyses in future work.

Several directions merit further exploration. First, integrating experimental data—such as bulk or single-cell transcriptomic profiles and perturbation assay results—into the current knowledge-driven framework may enhance the adaptability of inference results to specific biological contexts. Second, incorporating contextual annotations such as tissue type, cell type, or disease state may facilitate more accurate modeling of condition-dependent regulatory network rewiring. Finally, in clinical and precision medicine settings, applying this framework to patient-specific mutation combinations may provide valuable support for therapeutic target identification and resistance mechanism analysis.